# Reactive Force Field for P/Sn/I system: Atomistic Insight into the Early-Stage of Black Phosphorus and Phosphorene Synthesis Process

Djuric Brice Talonpa Tchoffo [a], Ismail Benabdallah* [a], Petr Neugebauer [b], Anouar Belhboub [c], Abdelouahad El Fatimy* [a]

[a] College of Physical Sciences and Engineering, Mohammed VI Polytechnic University, Ben Guerir, 43150, Morocco

[b] Central European Institute of Technology, CEITEC BUT, Purkynova 656/123, 61200 Brno, Czech Republic

[c] École Centrale Casablanca, Ville verte, Bouskoura, Casablanca, Morocco

## ABSTRACT:

Black phosphorus (BP) and its two-dimensional counterpart, phosphorene, are typically synthesized through chemical vapor transport (CVT) using Sn and $I_2$ additives. Chemical vapor deposition synthesis of phosphorene and allotropes is still yet not well understood. Investigating the atomistic mechanisms underlying phosphorus transport and early-stage processes is difficult experimentally. In this study, a reactive force field for the P/Sn/I system (P/Sn/I-2512) was developed and applied using ReaxFF-based molecular dynamics to explore the early-stage phase of the pre-nucleation relevant to BP/phosphorene growth. The force field parameters were trained on a comprehensive quantum-mechanical dataset covering bond dissociation, angle and torsion profiles, and tin condensed-phase equation-of-state and cluster formation energies, showing strong agreement in both gas and condensed phases. We demonstrate that iodine and density together control phosphorus recombination. Under low density, atomic phosphorus dominates with minimal clustering. Adding $I_2$ greatly increases P–P recombination, promotes the formation of $P_xI_y$ motifs, and transient $Sn_xP_yI_z$ compounds. Higher density systems favor the formation of larger $P_x$ clusters and support the development of ternary $Sn_xP_yI_z$ compounds that grow by capturing transported phosphorus. At the highest density, the system produces condensed, Hittorf-like phosphorus structures at the edges of $Sn_xP_yI_z$ clusters, along with BP-like hexagons stabilized by iodine that may act as nucleation seeds. These results offer an atomistic view of the transport and early-stage steps during the BP synthesis and provide a practical reactive model for studying growth conditions and additive effects in BP/phosphorene vapor synthesis.

## INTRODUCTION

Black phosphorus (BP), the most stable allotrope of phosphorus among red, white, and violet[1], was first discovered in 1914 by Bridgman[2]. Since its rediscovery in 2014 as a layered material composed of black phosphorene sheets by Liu et al. and Li et al.[3,4], black phosphorene has attracted significant attention due to its remarkable properties, including high carrier mobility (up to 1000 $cm^2/V\cdot s$), a high ON/OFF ratio (~$10^5$), and a tunable band gap ranging from 0.3 eV (bulk) to 2 eV (monolayer). It also exhibits a strong anisotropy[5,6]. These properties make BP a promising material that bridges the gap between graphene and transition-metal dichalcogenides[1].

Despite these advantages, the development of phosphorene-based devices remains limited by two major issues: their instability under ambient conditions[7] and the difficulty of producing large-area, high-quality black phosphorene films[8]. Common synthesis methods include bottom-up and top-down approaches. The bottom-up method remains incompletely understood due to the complex chemistry of phosphorus. In top-down methods, BP crystals are primarily used as raw materials for the synthesis of phosphorene. Therefore, the quality of the resulting phosphorene layer heavily depends on the quality of the BP used. Among bottom-up approaches, chemical vapor transport (CVT) has proven to be one of the most effective techniques for producing high-quality BP crystals under relatively moderate conditions[9–11]. In this process, red phosphorus is usually combined with tin (Sn) and iodine ($I_2$) to promote BP growth. While CVT allows for the formation of large crystals, issues such as secondary phases and limited control over the growth process persist.

Several studies have explored the mechanism of BP growth under CVT conditions to enhance both quality and yield. Key questions include: What role do additives play in BP growth? What are the most stable species at high temperature and pressure? Which species serve as nucleation sites? Are asked. Many experimental and theoretical studies have attempted to answer these questions, but clarity remains elusive. In 2016, Zhao et al. conducted experiments to examine the roles of metals and iodine in BP growth[11]. They found that iodine acts as a mineralizer, solubilizing phosphorus and supporting transport, while tin is crucial for crystallization. Other studies suggest that the final BP crystal forms via an intermediate called Hittorf's phosphorus (HP), and that the ternary clathrate ($Sn_{24}P_{19.3}I_8$) may act as a nucleation agent[12–14]. However, they concluded that the phase-transformation mechanism, with HP as a key intermediate, requires further refinement, as their evidence was based on low-purity by-products. The exact identity of the gas-phase species responsible for BP formation remains uncertain, and additional research is needed to clarify how red phosphorus transforms into BP[15]. In 2021, Pielmeier and Nilges argued that the presence of side phases near the target product does not necessarily indicate epitaxial growth[16]. They provided further insights through a combination of DFT calculations and thermodynamic analysis, suggesting that $P_4$ and $SnI_2$ are the dominant gas-phase species during phosphorene synthesis. Nonetheless, their conclusions rely on equilibrium-based approaches and reaction pathways modeled from structures optimized at 0K, making the formation mechanism largely speculative. Even with proposed mechanisms, alternative energetically favorable pathways might exist, highlighting the complexity of the reaction network and the limitations of static models. In 2023, Liu et al. attempted to develop a growth model based on the transformation mechanism, but they noted that their experiments, which involved either stopping synthesis midway or analyzing by-products directly, yielded unreliable results[17]. They faced challenges in experimentally understanding the growth process.

Computational methods offer a valuable alternative for investigating atomistic mechanisms that are difficult to access experimentally[18]. While density functional theory (DFT) provides high accuracy, it is limited to small systems and cannot simulate the dynamic environments involved[19]. In contrast, reactive molecular dynamics (ReaxFF-MD) can model large reactive systems at finite temperatures, making it especially useful for studying growth mechanisms and reaction pathways[18,19]. In chemical vapor deposition (CVD), reactions between precursors typically release species that are essential for the layer growth. As methane is used to release carbon atoms necessary for the growth of graphene, phosphorus atoms are released for the CVD growth of phosphorene. In this study, we simplified the process by taking the phosphorus atoms as a starting point to investigate the transport dynamics at high temperature during BP/phosphorene synthesis. Our work provides atomistic insights into the pre-nucleation stage of BP/phosphorene formation under highly reactive conditions. We develop and employ a ReaxFF reactive force field (P/Sn/I-2512) in molecular dynamics simulations to explore the behavior of the P/Sn/$I_2$ system and to identify a pre-nucleation pathway for early-stage BP/phosphorene formation, highlighting the important role of iodine in facilitating the recombination of P atoms.

# METHOD

## ReaxFF method

The ReaxFF reactive force field, first developed by van Duin et al[20], is a valuable tool for simulating complex reactive and physical systems. Unlike traditional non-reactive force fields[21,22], ReaxFF employs a bond-order-dependent approach to determine atom connectivity at each timestep during MD simulations. Because the bond order depends on distance, bond formation and breaking can occur dynamically throughout the simulation as the bond order decreases with increasing distance. The energy model in this framework includes various partial energy contributions[20,23]. Total energy is primarily divided into two parts: bond-order-dependent and non-bonded energy terms. Bond-order-dependent terms include bond dissociation, angle, and torsion contributions, while non-bonded terms account for van der Waals and Coulomb interactions. Non-bonded interactions are calculated between all atom pairs, regardless of their connectivity. Additionally, the atomic partial charges are computed using the self-consistent electronegativity equalization method (EEM), a geometry-dependent charge equilibration scheme[24].

The total potential energy of the system is expressed by the following equation:

$$E_{system} = E_{bond} + E_{val} + E_{over} + E_{under} + E_{tors} + E_{lp} + E_{vdw} + E_{coul} \quad (1)$$

where $E_{bond}$, $E_{val}$, $E_{over}$, $E_{under}$, $E_{tors}$, $E_{lp}$, $E_{vdw}$, and $E_{coul}$ represent bond formation and dissociation energy, three-body valence angle distortion energy, over-coordination penalty, under-coordination stabilization, four-body torsional angle energy, lone pair energy penalty, van der Waals interactions, and Coulomb interactions, respectively. A detailed description of each energy term can be found in van Duin's initial papers[20,23].

## Force Field development

The reference quantum-mechanical (QM) data used to train the force-field parameters were generated for both non-periodic and periodic molecular systems. Non-periodic QM calculations were performed using ADF/AMS 2025[25], employing the Perdew-Burke-Ernzerhof (PBE) functional from the generalized gradient approximation (GGA)[26] with a TZ2P basis set. Periodic QM calculations were performed using the BAND/AMS module[27] with the same GGA/PBE functional and the TZ2P basis set. All calculations included Grimme's empirical dispersion correction (D3)[28] to account for long-range van der Waals interactions. The convergence criteria for all calculations were set as follows: the energy change, the gradient, and the step convergence were set to $10^{-5}$ eV, 0.002 eV/Å, and $10^{-3}$ Å, respectively. Initial force field parameters were assembled by combining previously developed and validated parameters. Phosphorus (P) parameters were taken from the ReaxFF force field by Xiao et al.[29], while iodine

(I) parameters came from the force field reported by Pols et al.[30]. Lead (Pb) parameters from Pols et al.[30] were used as the starting point for tin (Sn) and were reoptimized during the parameter fitting. The QM training set was designed to cover a broad spectrum of chemical environments relevant to P–Sn–I systems, including both molecular and condensed-phase properties. QM training set included bond dissociation energies for Sn–Sn, Sn–I, Sn–P, I–P, and I–I bonds, as well as valence angle distortion profiles for configurations such as P–P–Sn, Sn–P–Sn, I–Sn–I, I–P–I, I–P–P, Sn–Sn–Sn, I–Sn–Sn, I–Sn–P, and I–P–Sn. Torsional energy scans were performed for these dihedral angles: I–Sn–Sn–I, I–P–P–I, I–Sn–P–P, I–Sn–Sn–Sn, and I–Sn–P–Sn. Additional data included reaction energies and heat of formation for key species like $SnI_2$, $SnI_4$, and $PI_3$ to ensure thermochemical consistency. To enhance transferability to larger systems, the training set was further extended with equations of state for $\alpha - Sn$, $\beta - Sn$, and formation energies of tin clusters ($Sn_x$, with x ranging from 2 to 6).

The following formula calculates the formation energy of the Sn clusters:

$$E_f = (E_{total} - x \times \mu_{Sn})/x \qquad (2)$$

Where $E_{total}$ denotes the total energy of a given cluster, x is the number of Sn atoms, and $\mu_{Sn}$ is the energy per Sn atom in bulk $\beta - Sn$.

## ReaxFF MD simulations methodology

Chemical vapor transport synthesis of BP typically involves three main phases: heating the system to generate vapor conditions, annealing at the target temperature (between 873 and 923K)[16] to enable transport with a minimal temperature gradient between the hot and cold zones, and finally cooling down. To examine the system's dynamics during the BP pre-nucleation phase (annealing), we used our P/Sn/I force field to perform reactive MD simulations with the open-source software LAMMPS[31].

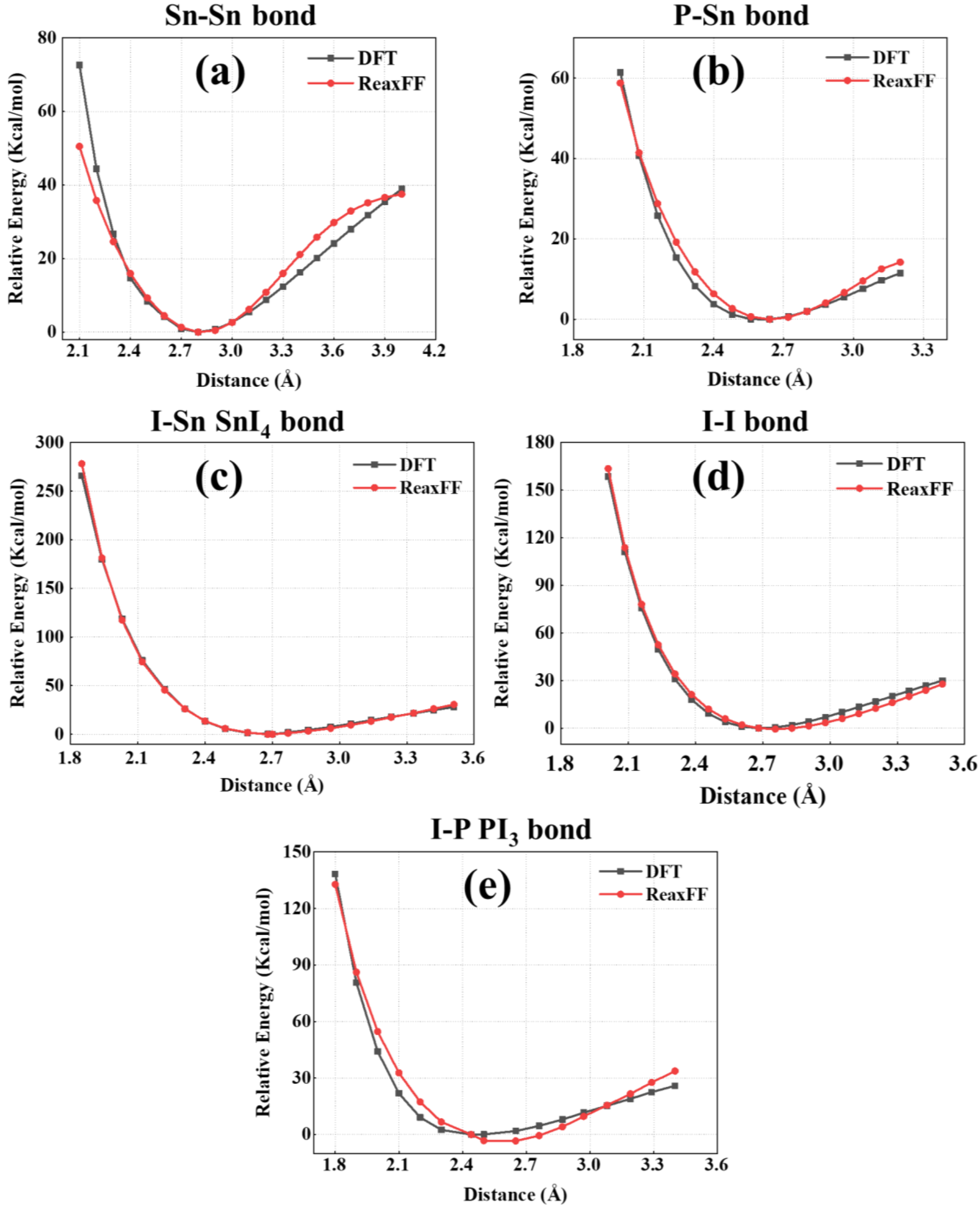

Fig. 1. Comparison between energy calculated from DFT (gray) and ReaxFF (red) for (a) Sn-Sn, (b) P-Sn, (c) I-Sn in $SnI_4$ molecule, (d) I-I, and (e) I-P pair interaction in $PI_3$.

First, to validate the formation of key species such as $SnI_2$, $SnI_4$, and $PI_3$ at high temperature (923 K), we perform multiple MD simulations on the Sn/$I_2$ and P/$I_2$ systems. The goal here is not to optimize the $Sn/I_2$ or $P/I_2$ reactions, but to confirm whether our force field produces the key species with correct geometry at high temperatures. For the $Sn/I_2$ reaction, we randomly place 30 Sn atoms and 30 $I_2$ molecules, representing a 1:1 $Sn/I_2$ ratio, in a cubic periodic box. For the $P/I_2$ reaction, we also place 30 P atoms and 45 $I_2$ molecules, representing a P/I2 ratio of 1:1.5, in a periodic cubic simulation box. In all our MD simulations, we use the NVT ensemble, which maintains the number of atoms (N), volume (V), and temperature (T) constant through a thermostat. The temperature is controlled with a Nosé-Hoover thermostat using a damping parameter of 100 fs for all simulations in this work. The timestep for the $Sn/I_2$ and $P/I_2$ reactions is 0.25 fs, and the total simulation time is 1 nanosecond (1 ns). Before running the 1 ns simulation, each system was energy-minimized. Because pressure influences the formation of species such as $SnI_4$ and $PI_3$, we evaluate its effect on each system ($Sn/I_2$ and $P/I_2$) by varying density while keeping it constant during the NVT simulation. The cubic simulation box lengths for each case study are 200 Å, 425 Å, and 910 Å, corresponding to effective pressures of 1 atm, 0.1 atm, and 0.01 atm at 923K, respectively.

To highlight the critical role of iodine in the recombination of P atoms, we used a larger simulation box (500 Å) corresponding to an effective pressure of 1 atm at 923K. Three different cases were studied. In the first case, the periodic cubic simulation box contains only 1000 P atoms. In the second case, we randomly placed 1000 P atoms and 50 Sn atoms in the simulation box; in the third case, we included 1000 P atoms, 50 Sn atoms, and 50 $I_2$ molecules to maintain a $P/Sn/I_2$ ratio of 20/1/1, which is commonly observed. The temperature was fixed at 923K, and the simulation time was set to 5 ns, which is sufficient to observe various phenomena.

Liu et al., in their study of the growth process of BP, demonstrated that when the pressure in the quartz tube was below 8 MPa, no BP was observed at the end of the synthesis. However, when the pressure exceeded 8 MPa, BP yield gradually increased[17]. Therefore, after highlighting the importance of iodine in the recombination of P atoms, the $P/Sn/I_2$ system was studied under three different densities: low (effective pressure of 5 atm), medium (effective pressure of 10 atm), high (effective pressure of 100 atm) corresponding to a cubic box of 300 Å, 240 Å, and 110 Å length, respectively. The same number of atoms was kept in all the simulations (1000 P, 50 Sn, and 50 $I_2$). The temperature (923 K), and total simulation time (5 ns) remained unchanged. The volume of the simulation box was calculated using the ideal-gas approximation at each effective pressure.

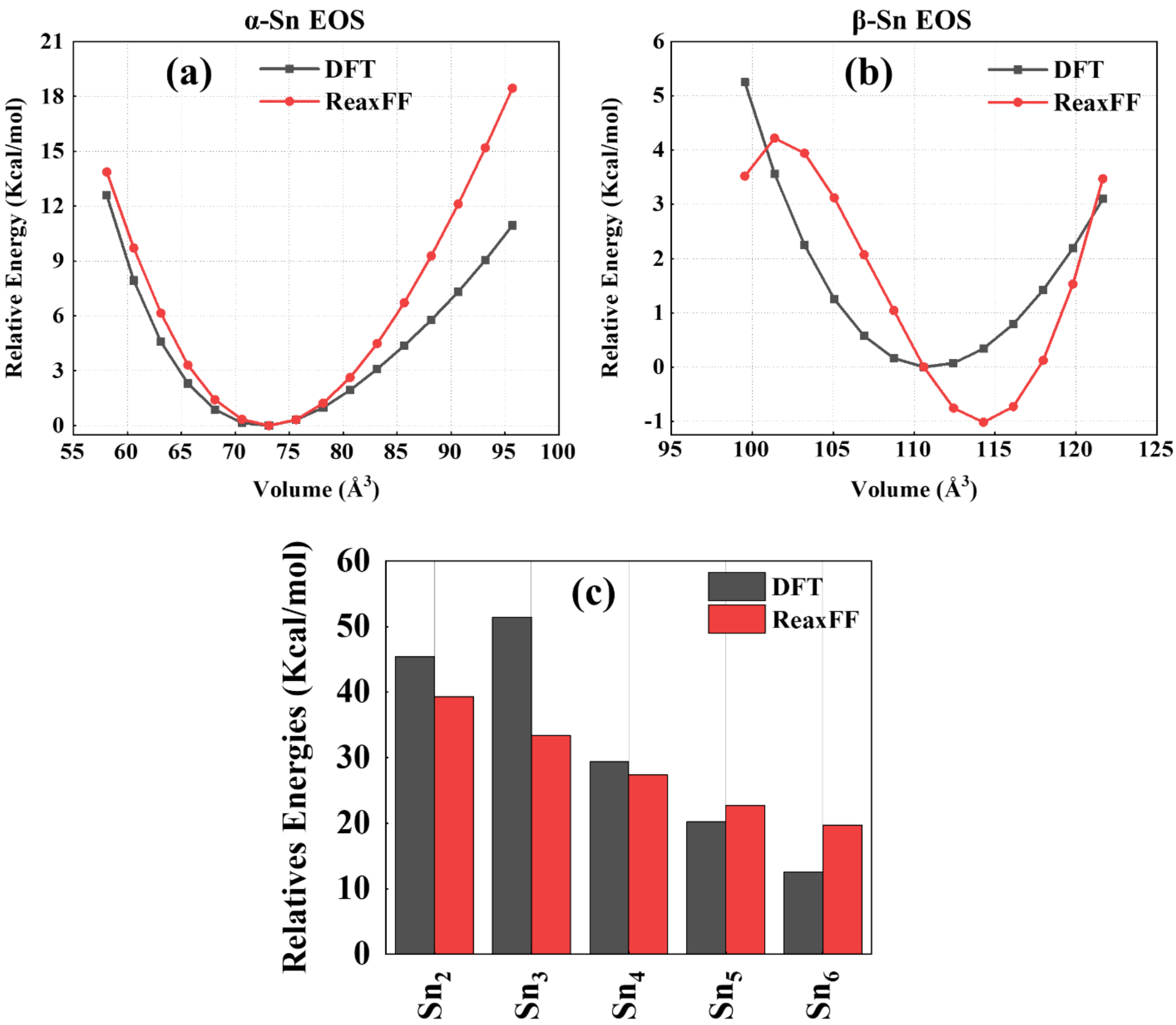

Fig. 2. Comparison between DFT (gray) and ReaxFF (red) equation of state (EOS) for (a) α-Sn, (b) β-Sn crystals, and (c) formation energies of $Sn_x$ clusters (x = 2 - 6).

These pressures should be interpreted as effective pressures used to define relative density conditions. At high densities where clustering and condensation occur, the ideal-gas approximation doesn't rigorously describe the thermodynamic state of the system. However, it provides a practical framework for systematically comparing the influence of density on clustering behavior. All simulations were performed within the NVT ensemble using a Nosé-Hoover thermostat with a damping parameter of 100 fs, and the timestep was set to 0.25 fs. The partial charge distribution was updated at each time step using the QEq scheme with a convergence tolerance of $10^{-6}$ and a spatial cutoff distance of 10 Å, thereby providing an accurate charge distribution during bond-breaking and bond-formation processes. System speciation at every 1000 timestep was determined using the "reaxff/species command" with a bond-order threshold of 0.3 for molecular identification[18]. This threshold was used solely for post-processing analysis and did not influence the ReaxFF simulation. To obtain statistically robust insights into the system's reactive behavior, 10 independent MD simulations were conducted, each initialized with a unique random configuration. The results were then averaged over all replicas.

# RESULTS AND DISCUSSION

## A. Force field development

The ReaxFF parameter set for the P/Sn/I system was trained against quantum mechanical (QM) reference data using the Covariance Matrix Adaptation Evolutionary Strategy (CMA-ES)[32], as implemented in the ParAMS module[33] within the Amsterdam Modeling Suite (AMS 2025)[25]. This global optimization method is especially effective for tackling the highly coupled and nonlinear parameter space typical of reactive force fields. The training set used to optimize the force field includes various quantities, such as bond dissociation energies, valence angle bending energies, torsion energies, charges, heats of formation, and tin crystal equations of state. As mentioned earlier, the phosphorus-related force field parameters used in the present work were inherited from the previously validated ReaxFF parametrization of Xiao et al., which already demonstrated accurate reproduction of structural, thermodynamic, and mechanical properties of bulk black phosphorus, black phosphorene, and blue phosphorene.[29]

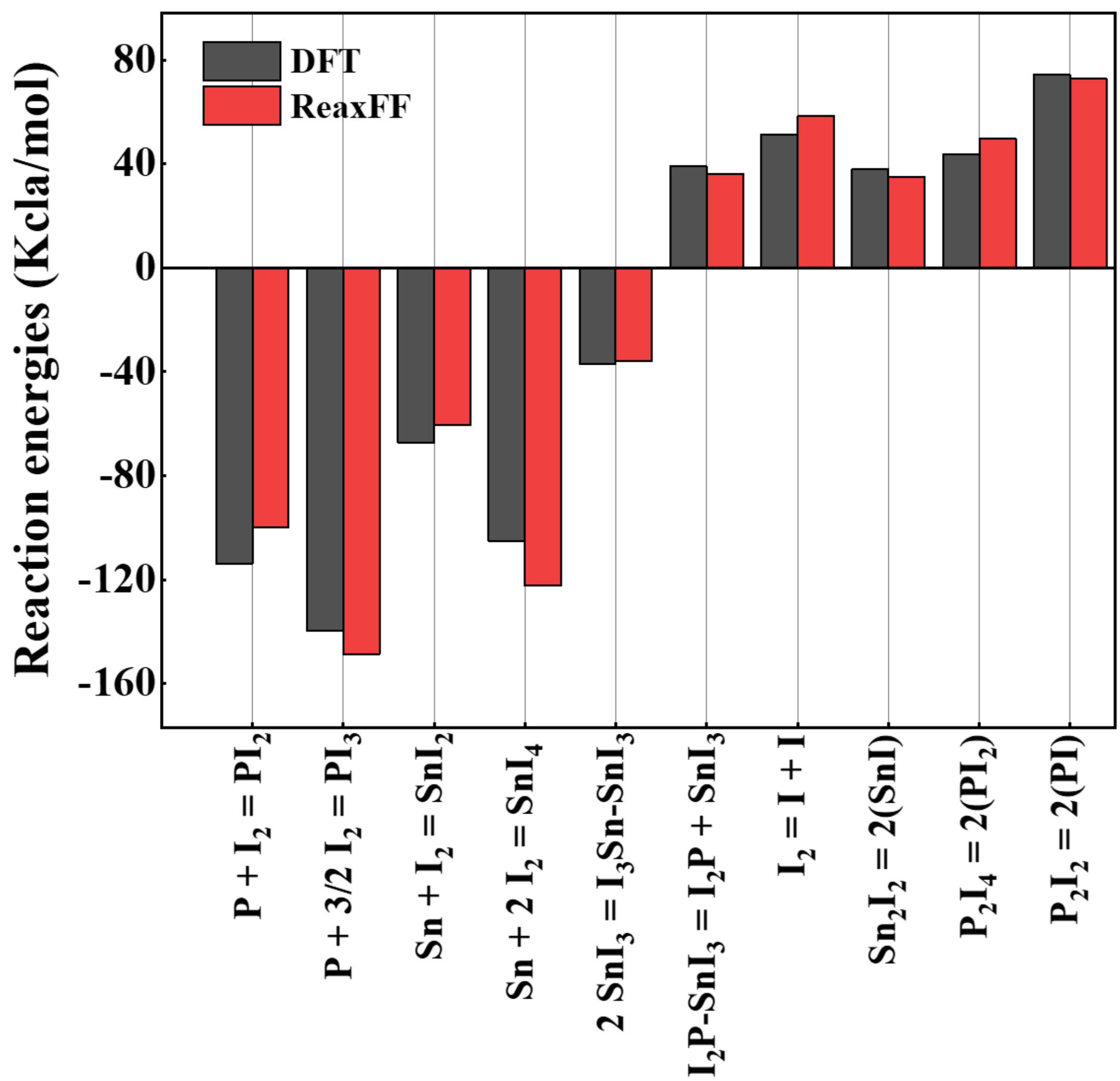

Fig. 3. Comparison between reaction energies obtained with DFT (gray) and ReaxFF (red).

**Figure 1** compares ReaxFF and QM results for each pair interaction. (a) ReaxFF vs DFT for Sn-Sn interaction, (b) P-Sn interaction, (c) I-Sn interaction, (d) I-I interaction, and (e) I-P interaction pairs. The P-P dissociation energy was previously determined by Xiao et al. in the development of the P/H reactive force field[29]. This figure shows that the optimized force field accurately reproduces the DFT bond dissociation energies, providing a reliable basis for MD simulations of the P/Sn/I chemical system. Valence and torsional angles are also well represented by the force field parameters. This is confirmed by the good agreement between ReaxFF and QM reference data for valence energies and dihedral rotational barriers (Figures S2, S3, and S4).

As mentioned in the methods section, the initial Sn description used lead (Pb) parameters from the force field developed by Pots *et al.*[30]. Subsequently, these parameters were reoptimized to accurately describe tin chemistry and condensed-phase. In particular, the fitting focused on the equations of state $\alpha - Sn$, $\beta - Sn$, and formation energies of Sn clusters. As shown in **Figure 2**, ReaxFF reproduces the corresponding QM reference data with good agreement. **Figure 3** shows the comparison between DFT and ReaxFF for the reaction energies of important reactions, such as the formation of $SnI_2$, $SnI_4$, $PI_2$, $PI_3$, etc. The ReaxFF energy descriptions are in good agreement with QM data, demonstrating the force field's ability to accurately describe the dynamics of the P/Sn/I system. The Force field parameters can be found in the supplementary material.

## B. Molecular Dynamics Simulations of Sn/$I_2$ and P/$I_2$ systems at 923 K (density effect)

Numerous studies have shown that synthesizing pure $SnI_2$ is difficult due to the constant competition with $SnI_4$[34,35]. To demonstrate our force field's ability to accurately describe the P/Sn/I system, we performed a series of molecular dynamics simulations of the Sn/$I_2$ system at 923 K under different effective pressures. The temperature of 923 K was chosen to match other experimental studies[16,17]. As explained in the methods section, the box size was adjusted to change the system density by keeping the number of particles constant. **Figure 4a** illustrates the species formation dynamics at effective pressures of 0.01 atm, 0.1 atm, and 1 atm. The use of a $Sn/I_2$ ratio of 1:1 naturally favors the formation of $SnI_2$. The simulations show that density influences the chemical composition of the $Sn/I_2$ system. As the density increases, the system gradually shift from a dilute state dominated by elemental Sn and $I_2$ to more complex Sn-I coordination environments, including simple tin iodide compounds (SnI, $SnI_2$, $Sn_2I_2$) at 0.1 atm, and extended Sn-I aggregates ($SnI_4$, $Sn_2I_4$, etc.) at 1 atm. The simulations indicate that $SnI_2$ remains the main species across all effective pressures, but higher density promotes the formation of larger Sn-I molecules, such as $SnI_4$, due to increased interactions between $SnI_2$ and other molecules, including $I_2$, despite their limited abundance. This is significant because the competition between Sn(II) and Sn(IV) iodide species is a known challenge in tin-iodide chemistry, occurring in both bulk synthesis and vapor-phase processes where partial iodination of $SnI_2$ to $SnI_4$ is common[36]. An additional phase observed at higher pressures is $Sn_2I_4$, which Pielmeier et al. reported forming through dimerization of two $SnI_2$ molecules in a saturated environment, exactly what we observe at high density, as shown by the $SnI_2$ formation curve at an effective pressure of 1 atm[16]. They identified $Sn_2I_4$ as a key species in the formation of the 1D SnPI chain. We also performed multiple MD simulations to study the reaction between phosphorus and diiodine. 30 P atoms were combined with 45 $I_2$ molecules in a simulation box, whose volume was adjusted to the target effective pressures (0.01 atm, 0.1 atm, and 1 atm) for NVT simulations. **Figure 4b** shows the formation of various species during the simulation at 923 K under different system densities. At very low density (effective pressure of 0.01 atm), the system remains mostly atomic, with only a small amount of $PI_2$ forming. Lower density limits interactions between species, making reactions less likely. When the density increased corresponding to an effective pressure of 0.1 atm, the system became more reactive, producing more $PI_2$ and some $PI_3$, which results from interactions between PI and $I_2$. At 1 atm, the formation of $PI_2$ accelerates, and more $PI_3$ appears, but $PI_2$ still dominates. The coexistence of $PI_3$ with other $PI_x$ phases aligns with experimental results, which show that isolating pure $PI_3$ is challenging because it often associates with mixed iodide species and polymeric structures[37].

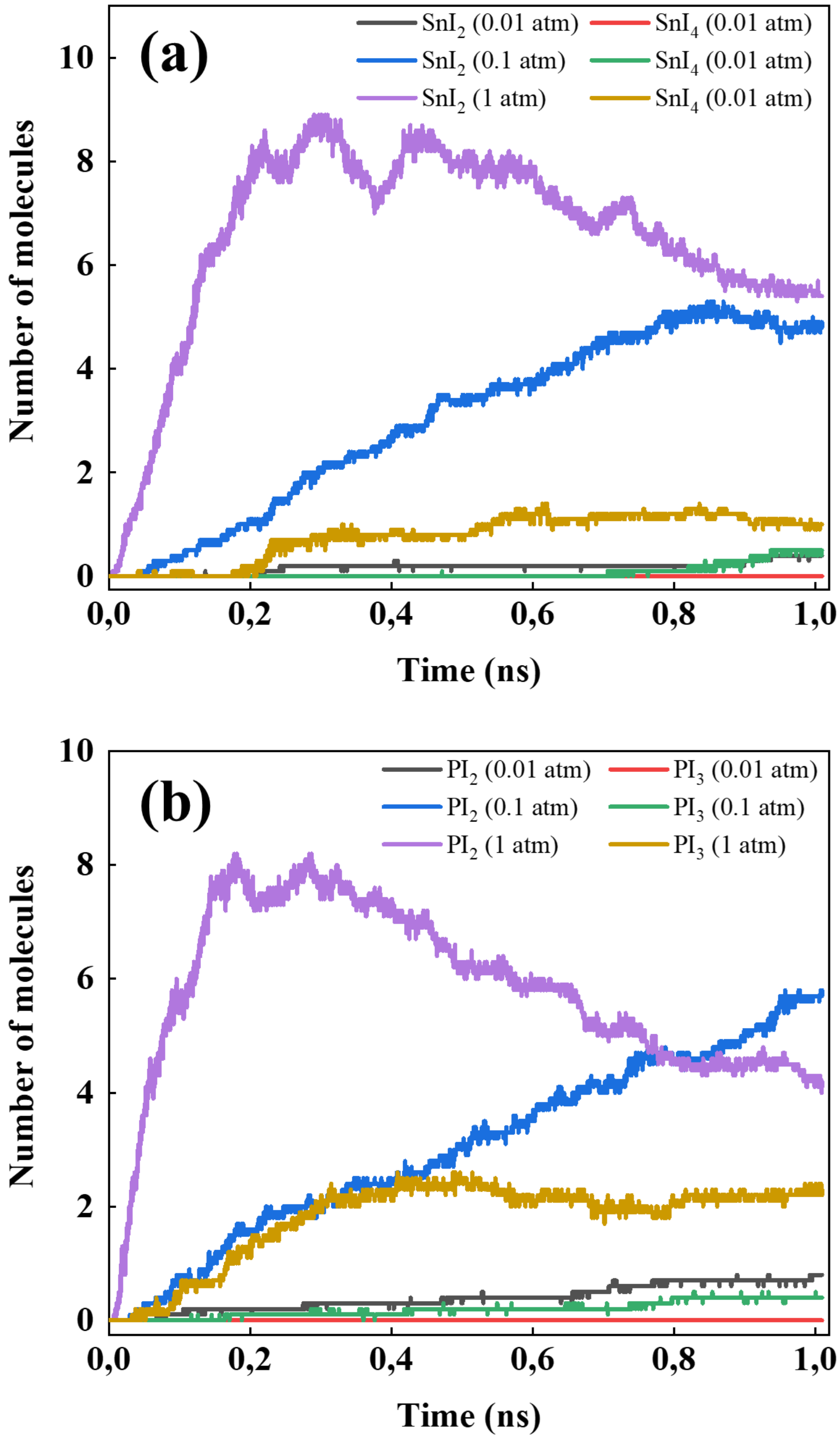

Fig. 4. (a) Formation of $SnI_2$ and $SnI_4$ molecules in the Sn/$I_2$ system, (b) Formation of $PI_2$ and $PI_3$ in the P/$I_2$ system at different effective pressures.

### C. Role of iodine in the recombination process of phosphorus atoms

To understand the growth mechanism of a material, the role of each element should be specified. Our system consists of phosphorus, tin, and diiodine. To investigate iodine's role in the recombination of $P_x$ clusters, we performed reactive MD simulations for three different systems under the same conditions. The first system contains a box with 1000 P atoms; the second contains a box with 1000 P atoms and 50 Sn atoms; and the third includes 1000 P atoms, 50 Sn atoms, and 50 $I_2$ molecules, maintaining a 1:1 ratio between Sn and $I_2$. The temperature was set to 923 K, and the systems were subjected to an effective pressure of 1 atm with a timestep of 0.25 fs. The total simulation time was 5 nanoseconds (ns). The number of Px clusters was tracked every 25 ps (1000 timesteps) during the simulation, as shown in **Figure 5**. This figure indicates that at this effective pressure, the system consists predominantly of atoms. However, we observe the evolution of the number of $P_2$ clusters across all simulations over time. A higher number of $P_2$ recombination events from P atoms occurs when $I_2$ is present, as evidenced by the red curve. In contrast, the dynamics of $P_2$ recombination in the case of only P atoms (gray curve) and the mixed P and Sn systems (blue curve) are quite similar. Experimental studies, starting with red phosphorus in the Sn-I-assisted synthesis of BP, have demonstrated that iodine is crucial for solubilizing phosphorus and facilitating its transport from the hot to the cold zone[11]. Additionally, analysis of the system at high temperature revealed the coexistence of $P_4$ and $P_2$ molecules, even though $P_4$ dominates[16,17]. These simulations highlight iodine's role in recombination and phosphorus transport, as indicated by the presence of binary $P_xI_y$ and ternary PSnI motifs in the system. This P-to-$P_2$ recombination phenomenon, observed near atmospheric pressure in the presence of iodine, will be very important for optimizing the growth time of the phosphorene allotrope synthesis in the atmospheric-pressure CVD framework.

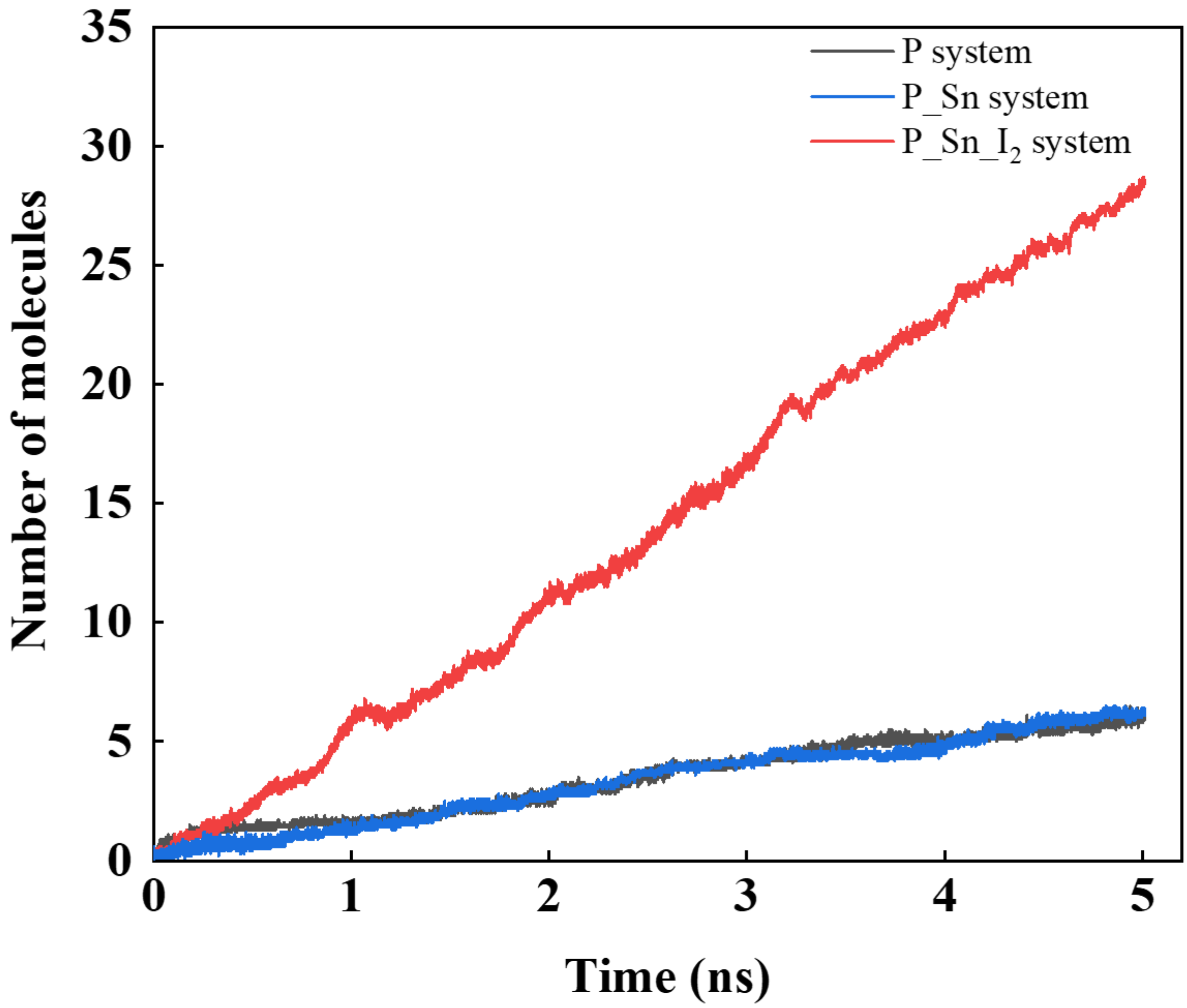

Fig. 5. Dynamics of formation of phosphorus $P_2$ cluster in a system containing only phosphorus (gray), phosphorus with tin (blue), and finally phosphorus with tin and iodine (red) during the simulation.

### D. Density study and prenucleation formation mechanism

Recently, Liu et al. studied the effect of pressure on the synthesis of BP by varying the length of the quartz tube in which the synthesis occurs[17]. They showed that no BP formed at pressures below 8 MPa, but when the pressure exceeded 8 MPa, the material formed. To further examine the density effect, we also performed a series of MD simulations. The same system composition (1000 P, 50 Sn, and 50 $I_2$) was used for all simulations. The temperature remained constant at 923 K, while three different system densities were studied, corresponding to effective pressures of 5 atm, 10 atm and 100 atm. The ideal-gas approximation was used to estimate the simulation box dimensions corresponding to the three relative density regimes. So, the reported effective pressures should not be interpreted as exact pressures of the system. **Figures 6 (a)** to **(c)** illustrate the formation dynamics of various phosphorus clusters. At low density (effective pressure of 5 atm), the system mainly consists of phosphorus atoms (**Figure S6**), $P_2$, $P_3$, and a small amount of $P_4$ molecules, as shown in **Figure 6 (a)**. **Figure 6 (b)** indicates that at medium density (effective pressure of 10 atm), larger phosphorus clusters begin to appear, with $P_2$ remaining predominant. By the end of the simulation, some P atoms still exist in the system, as depicted in **Figure S6.** The decline in the $P_2$ curves (gray) at the end of the simulation reflects the transformation of some $P_2$ molecules into higher-order molecules, as evidenced by increases in $P_4$ (blue) and $P_5$ (green) molecules. **Figure 6 (c)** shows that at high density (effective pressure of 100 atm), larger molecules form more readily. Compared to lower densities, the cluster formation rate at this high density is significantly higher, highlighting the role of the high-density regimes for phosphorus clustering and prenucleation. By the end of the simulation, no P atoms remain in the system, as shown in **Figure S6**. The only detectable clusters are high-order $P_x$ (x = 8 - 12). During the simulation, many other species were observed, such as binary $P_xI_y$ motifs and ternary $Sn_xP_yI_z$ clusters. The presence of these $P_xI_y$ motifs supports Zhao et al.'s claim, emphasizing iodine's key role in phosphorus transport. The size of these clusters increases with density, confirming that higher density promotes the formation of larger clusters. This behavior may be relevant to BP formation, as Liu et al. demonstrated that low-density systems consisting of light species do not permit BP formation, whereas high-density systems lead to more condensed phases, shown by the high-order $Sn_xP_yI_z$ clusters present at the end of the simulation. **Figure 6 (d)** provides snapshots of one such large $Sn_xP_yI_z$ cluster at the end of the simulation. Based on the evolution of the high-density P/Sn/$I_2$ system, the observed prenucleation pathway can be described as follows: in the early reaction stages, the combined effect of iodine and high-density conditions promotes rapid recombination of P atoms into $P_x$ small clusters and the formation of small $Sn_xP_yI_z$ clusters. As the reaction progresses, these clusters grow by adsorbing P atoms, transported via iodine and $P_x$ clusters. By the end, the edges of the $Sn_xP_yI_z$ cluster, as shown in **Figure 6 (e)**, are surrounded by violet phosphorus-like clusters. These findings are consistent with experimental studies suggesting the involvement of Hittorf-like phosphorus as an intermediate phase during BP formation, with $Sn_{24}P_{19.3}I_8$ acting as the nucleation agent[13,14,17]. Importantly, the simulation reveals a BP-like hexagonal motif that may represent a potential nucleation seed (**Figure S7**). These species may serve as potential precursors for BP growth during the slow cooling stage.

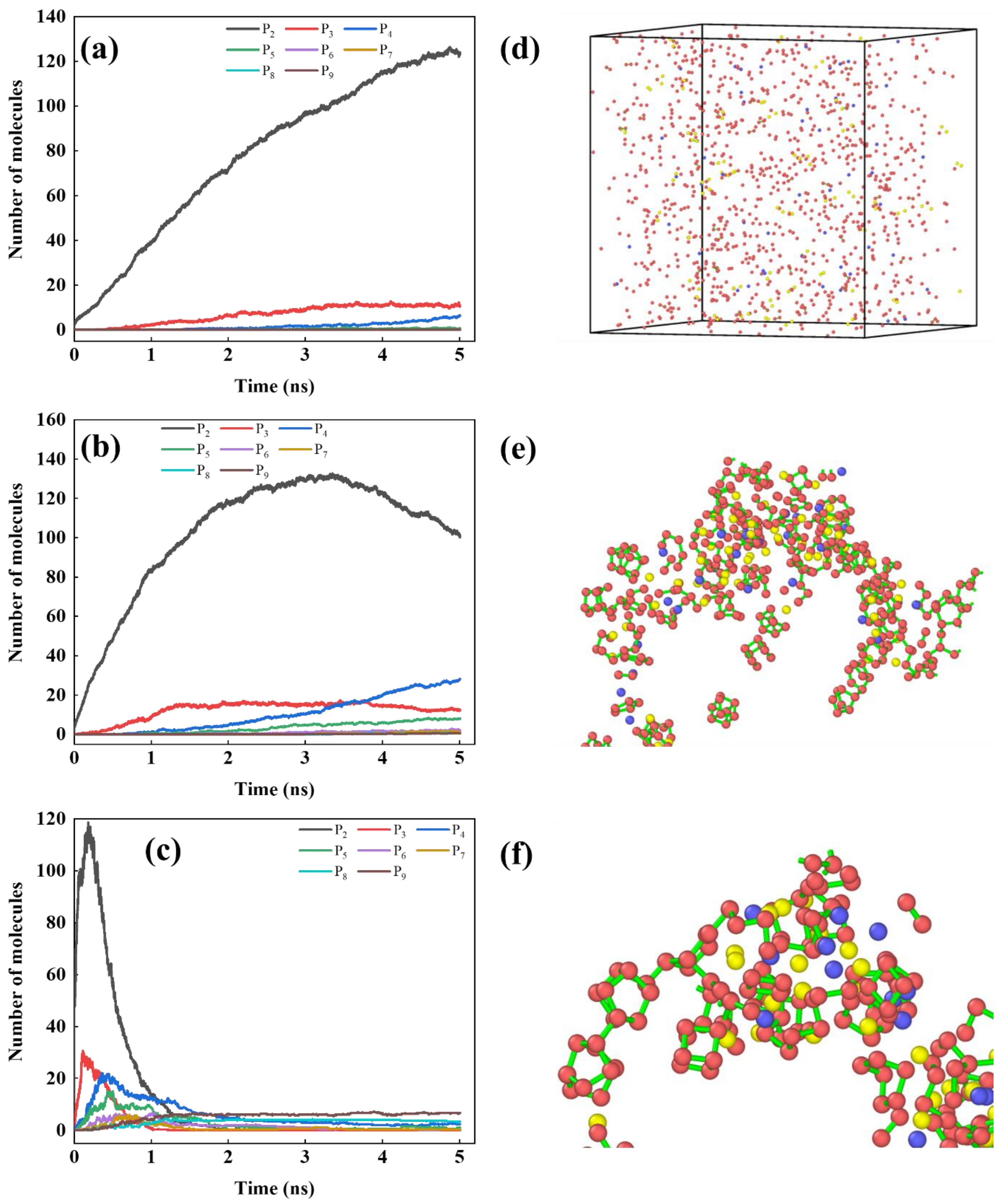

Fig. 6. Dynamics of formation of phosphorus clusters in P/Sn/$I_2$ system under (a) Low density, (b) Medium density, (c) High density. (d) Snapshot of the equilibrated P/Sn/$I_2$ system at 923K under high density (e). Snapshot of $Sn_xP_yI_z$ ternary compound at the end of the simulation of the P/Sn/$I_2$ system under high density (f). Snapshot of the Hittorf-like phosphorus cluster at the edge of the $Sn_xP_yI_z$ compound. Red atoms, blue atoms, and yellow atoms are phosphorus, tin, and iodine, respectively.

## CONCLUSION

In summary, we performed reactive molecular dynamics simulations to investigate the early-stage formation process of BP/phosphorene under various conditions and to emphasize the important role of iodine in the recombination of phosphorus atoms. We developed a ReaxFF reactive force field for the P/Sn/I system, trained against extensive QM data. This force field demonstrates a good agreement with DFT calculations. Further validation was achieved through reactive MD simulations of the Sn/$I_2$ and P/$I_2$ systems at 923 K under different densities. Studying the dynamics of these systems reveals a pressure dependence in the formation of key species, such as $SnI_2$, $SnI_4$, and $Sn_2I_4$ in the Sn/$I_2$ system, and $PI_2$, $PI_3$, and $P_xI_y$ in the P/$I_2$ system. The results also demonstrate the essential role of iodine in improving phosphorus solubility, transport, and recombination, evidenced by the increased number of phosphorus clusters and the strong reactions between phosphorus and iodine, which form not only $P_xI_y$ motifs but also serve as links in ternary PSnI compounds. Our findings suggest that high-density conditions promote the prenucleation of BP/phosphorene by condensing phosphorus atoms into Hittorf-like phosphorus, an intermediate step toward BP formation. Finally, analyzing the system's dynamics allows us to propose this prenucleation pathway: initially, the combined action of iodine and increasing density facilitates the recombination of isolated phosphorus atoms and the formation of ternary $Sn_xP_yI_z$ compounds. These $Sn_xP_yI_z$ compounds then grow by adsorbing phosphorus clusters and phosphorus atoms transported via iodine, with excess phosphorus atoms condensing into HP-like phosphorus at the edges of the $Sn_xP_yI_z$ compounds. Additionally, potential BP nucleation seeds were identified as BP-like hexagons stabilized by iodine atoms. These findings enhance our understanding of the black phosphorus/phosphorene early-stage growth process and provide valuable insights into the role of each precursor.

**REFERENCE**:

(1) Chaudhary, V.; Neugebauer, P.; Mounkachi, O.; Lahbabi, S.; El Fatimy, A. Phosphorene—an Emerging Two-Dimensional Material: Recent Advances in Synthesis, Functionalization, and Applications. *2D Mater.* **2022**, *9* (3), 032001.

(2) Bridgman, P. Two New Modifications of Phosphorus. *J. Am. Chem. Soc.* **1914**, *36* (7), 1344–1363.

(3) Liu, H.; Neal, A. T.; Zhu, Z.; Luo, Z.; Xu, X.; Tománek, D.; Ye, P. D. Phosphorene: An Unexplored 2D Semiconductor with a High Hole Mobility. *ACS Nano* **2014**, *8* (4), 4033–4041.

(4) Li, L.; Yu, Y.; Ye, G. J.; Ge, Q.; Ou, X.; Wu, H.; Feng, D.; Chen, X. H.; Zhang, Y. Black Phosphorus Field-Effect Transistors. *Nat. Nanotechnol.* **2014**, *9* (5), 372–377.

(5) Woomer, A. H.; Farnsworth, T. W.; Hu, J.; Wells, R. A.; Donley, C. L.; Warren, S. C. Phosphorene: Synthesis, Scale-up, and Quantitative Optical Spectroscopy. *ACS Nano* **2015**, *9* (9), 8869–8884.

(6) Jing, Y.; Tang, Q.; He, P.; Zhou, Z.; Shen, P. Small Molecules Make Big Differences: Molecular Doping Effects on Electronic and Optical Properties of Phosphorene. *Nanotechnology* **2015**, *26* (9), 095201.

(7) Favron, A.; Gaufrès, E.; Fossard, F.; Phaneuf-L'Heureux, A.-L.; Tang, N. Y.; Lévesque, P. L.; Loiseau, A.; Leonelli, R.; Francoeur, S.; Martel, R. Photooxidation and Quantum Confinement Effects in Exfoliated Black Phosphorus. *Nat. Mater.* **2015**, *14* (8), 826–832.

(8) Castellanos-Gomez, A. Black Phosphorus: Narrow Gap, Wide Applications. *J. Phys. Chem. Lett.* **2015**, *6* (21), 4280–4291.

(9) Lange, S.; Schmidt, P.; Nilges, T. Au3SnP7@ Black Phosphorus: An Easy Access to Black Phosphorus. *Inorg. Chem.* **2007**, *46* (10), 4028–4035.

(10) Köpf, M.; Eckstein, N.; Pfister, D.; Grotz, C.; Krüger, I.; Greiwe, M.; Hansen, T.; Kohlmann, H.; Nilges, T. Access and in Situ Growth of Phosphorene-Precursor Black Phosphorus. *J. Cryst. Growth* **2014**, *405*, 6–10.

(11) Zhao, M.; Niu, X.; Guan, L.; Qian, H.; Wang, W.; Sha, J.; Wang, Y. Understanding the Growth of Black Phosphorus Crystals. *CrystEngComm* **2016**, *18* (40), 7737–7744.

(12) Zhang, Z.; Xing, D.-H.; Li, J.; Yan, Q. Hittorf's Phosphorus: The Missing Link during Transformation of Red Phosphorus to Black Phosphorus. *CrystEngComm* **2017**, *19* (6), 905–909.

(13) Wang, D.; Yi, P.; Wang, L.; Zhang, L.; Li, H.; Lu, M.; Xie, X.; Huang, L.; Huang, W. Revisiting the Growth of Black Phosphorus in Sn-I Assisted Reactions. *Front. Chem.* **2019**, *7*, 21.

(14) Chen, Z.; Zhu, Y.; Lei, J.; Liu, W.; Xu, Y.; Feng, P. A Stage-by-Stage Phase-Induction and Nucleation of Black Phosphorus from Red Phosphorus under Low-Pressure Mineralization. *CrystEngComm* **2017**, *19* (47), 7207–7212.

(15) Khurram, M.; Sun, Z.; Zhang, Z.; Yan, Q. Chemical Vapor Transport Growth of Bulk Black Phosphorus Single Crystals. *Inorg. Chem. Front.* **2020**, *7* (15), 2867–2879.

(16) Pielmeier, M. R.; Nilges, T. Formation Mechanisms for Phosphorene and SnIP. *Angew. Chem. Int. Ed.* **2021**, *60* (12), 6816–6823.

(17) Liu, W.; Zhu, Y.; Chen, S.; Ban, X.; Gu, X.; Hu, F.; Wang, L.; Du, W. Black-Phosphorus Crystal Growth Model Deduced from the Product Distribution State under Different Process Factors. *CrystEngComm* **2023**, *25* (31), 4470–4479.

(18) Ashraf, C.; Van Duin, A. C. Extension of the ReaxFF Combustion Force Field toward Syngas Combustion and Initial Oxidation Kinetics. *J. Phys. Chem. A* **2017**, *121* (5), 1051–1068.

(19) Uene, N.; Mabuchi, T.; Zaitsu, M.; Yasuhara, S.; van Duin, A. C.; Tokumasu, T. Reactive Force Field Molecular Dynamics Studies of the Initial Growth of Boron Nitride Using BCl3 and NH3 by Atomic Layer Deposition. *J. Phys. Chem. C* **2024**, *128* (3), 1075–1086.
(20) Van Duin, A. C.; Dasgupta, S.; Lorant, F.; Goddard, W. A. ReaxFF: A Reactive Force Field for Hydrocarbons. *J. Phys. Chem. A* **2001**, *105* (41), 9396–9409.
(21) Weiner, P. K.; Kollman, P. A. AMBER: Assisted Model Building with Energy Refinement. A General Program for Modeling Molecules and Their Interactions. *J. Comput. Chem.* **1981**, *2* (3), 287–303.
(22) Brooks, B. R.; Brooks III, C. L.; Mackerell Jr, A. D.; Nilsson, L.; Petrella, R. J.; Roux, B.; Won, Y.; Archontis, G.; Bartels, C.; Boresch, S. CHARMM: The Biomolecular Simulation Program. *J. Comput. Chem.* **2009**, *30* (10), 1545–1614.
(23) Van Duin, A. C.; Strachan, A.; Stewman, S.; Zhang, Q.; Xu, X.; Goddard, W. A. ReaxFFSiO Reactive Force Field for Silicon and Silicon Oxide Systems. *J. Phys. Chem. A* **2003**, *107* (19), 3803–3811.
(24) Mortier, W. J.; Ghosh, S. K.; Shankar, S. Electronegativity-Equalization Method for the Calculation of Atomic Charges in Molecules. *J. Am. Chem. Soc.* **1986**, *108* (15), 4315–4320.
(25) Baerends, E. J.; Aguirre, N. F.; Austin, N. D.; Autschbach, J.; Bickelhaupt, F. M.; Bulo, R.; Cappelli, C.; van Duin, A. C.; Egidi, F.; Fonseca Guerra, C. The Amsterdam Modeling Suite. *J. Chem. Phys.* **2025**, *162* (16).
(26) Perdew, J. P. Generalized Gradient Approximation Made Simple. *Phys Rev Lett* **1997**, *77*, 3868.
(27) Te Velde, G.; Baerends, E. Precise Density-Functional Method for Periodic Structures. *Phys. Rev. B* **1991**, *44* (15), 7888.
(28) Grimme, S.; Antony, J.; Ehrlich, S.; Krieg, H. A Consistent and Accurate Ab Initio Parametrization of Density Functional Dispersion Correction (DFT-D) for the 94 Elements H-Pu. *J. Chem. Phys.* **2010**, *132* (15).
(29) Xiao, H.; Shi, X.; Hao, F.; Liao, X.; Zhang, Y.; Chen, X. Development of a Transferable Reactive Force Field of P/H Systems: Application to the Chemical and Mechanical Properties of Phosphorene. *J. Phys. Chem. A* **2017**, *121* (32), 6135–6149.
(30) Pols, M.; Vicent-Luna, J. M.; Filot, I.; Van Duin, A. C.; Tao, S. Atomistic Insights into the Degradation of Inorganic Halide Perovskite Cspbi3: A Reactive Force Field Molecular Dynamics Study. *J. Phys. Chem. Lett.* **2021**, *12* (23), 5519–5525.
(31) Plimpton, S. Fast Parallel Algorithms for Short-Range Molecular Dynamics. *J. Comput. Phys.* **1995**, *117* (1), 1–19.
(32) Hansen, N.; Ostermeier, A. Completely Derandomized Self-Adaptation in Evolution Strategies. *Evol. Comput.* **2001**, *9* (2), 159–195.
(33) Komissarov, L.; Ruger, R.; Hellstrom, M.; Verstraelen, T. ParAMS: Parameter Optimization for Atomistic and Molecular Simulations. *J. Chem. Inf. Model.* **2021**, *61* (8), 3737–3743.
(34) Zuraw, W.; Kubicki, D.; Kudrawiec, R.; Przypis, Ł. Carboxylic Acid-Assisted Synthesis of Tin (II) Iodide: Key for Stable Large-Area Lead-Free Perovskite Solar Cells. *ACS Energy Lett.* **2024**, *9* (9), 4509–4515.
(35) Jiang, X.; Li, H.; Zhou, Q.; Wei, Q.; Wei, M.; Jiang, L.; Wang, Z.; Peng, Z.; Wang, F.; Zang, Z. One-Step Synthesis of $SnI_2 \cdot (DMSO)_x$ Adducts for High-Performance Tin Perovskite Solar Cells. **2021**.
(36) Brekhovskikh, M.; Mastryukov, M.; Kornev, P.; Gasanov, A.; Kovalenko, A.; Fedorov, V. Synthesis and Ultrapurification of Tin Diiodide. *Inorg. Mater.* **2019**, *55* (9), 974–978.
(37) Brauer, G. *Handbook of Preparative Inorganic Chemistry V2*; Elsevier, 2012; Vol. 2.

(38) Kissami, I., Basmadjian, R., Chakir, O., & Abid, M. R. TOUBKAL: a high-performance supercomputer powering scientific research in Africa: I. Kissami et al. *The Journal of Supercomputing*. **2025** *81*(15), 1401.

# Supporting Material

## Reactive Force Field for P/Sn/I system: Atomistic Insight into the Early-Stage of Black Phosphorus and Phosphorene Synthesis Process

Djuric Brice Talonpa Tchoffo [a], Ismail Benabdallah* [a], Petr Neugebauer [b], Anouar Belhboub[c], Abdelouahad El Fatimy* [a]

[a] College of Physical Sciences and Engineering, Mohammed VI Polytechnic University, Ben Guerir, 43150, Morocco

[b] Central European Institute of Technology, CEITEC BUT, Purkynova 656/123, 61200 Brno, Czech Republic

[c] École Centrale Casablanca, Ville verte, Bouskoura, Casablanca, Morocco

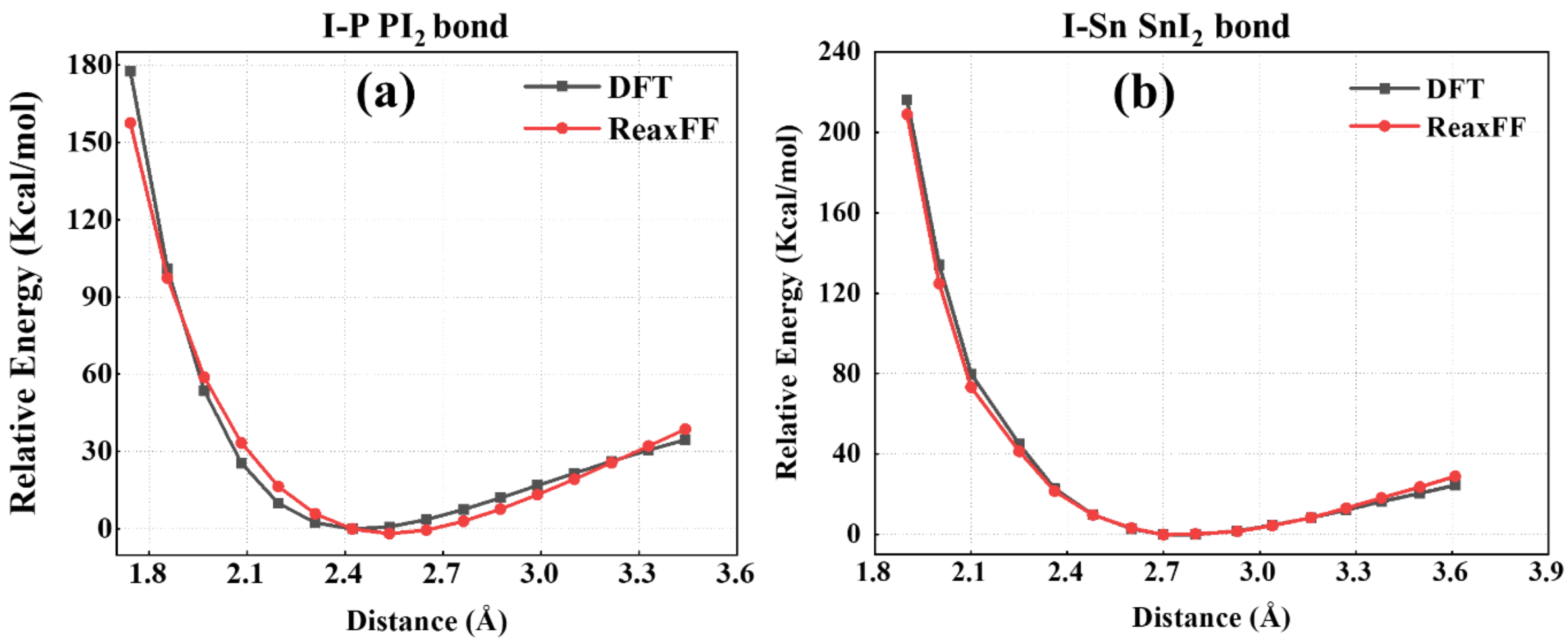

Fig. S1: Comparison between energy calculated from DFT (gray) and ReaxFF (red) for (a) I-P pair interaction in $PI_2$ and (b) I-Sn in $SnI_2$ molecule.

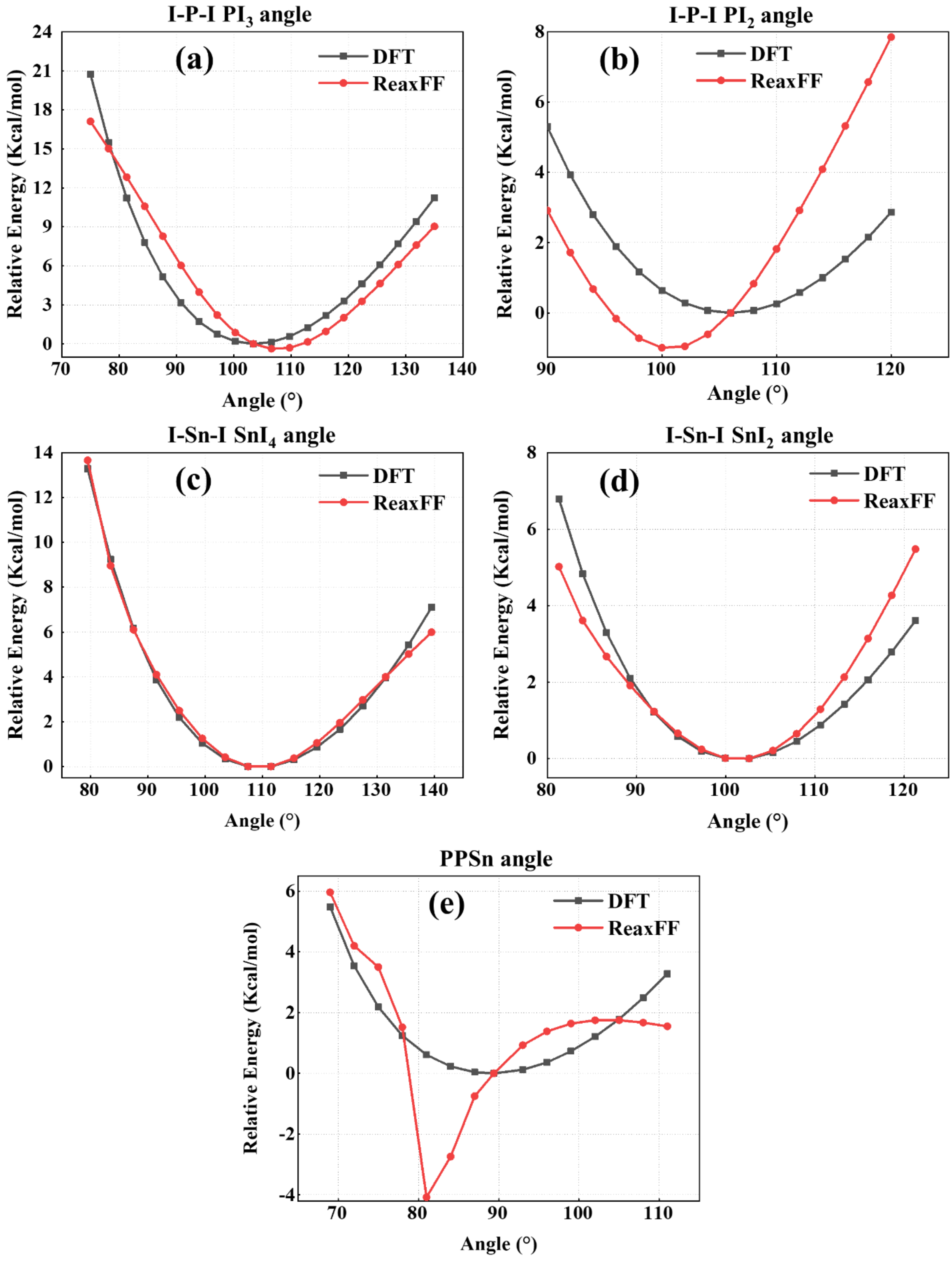

Fig. S2: Comparison between energy calculated from DFT (gray) and ReaxFF (red) for valence angle distortion (a) I-P-I in $PI_3$ (b) I-P-I in $PI_2$ (c) I-Sn-I in $SnI_4$ (c) I-Sn-I in $SnI_2$ (c) P-P-Sn angles.

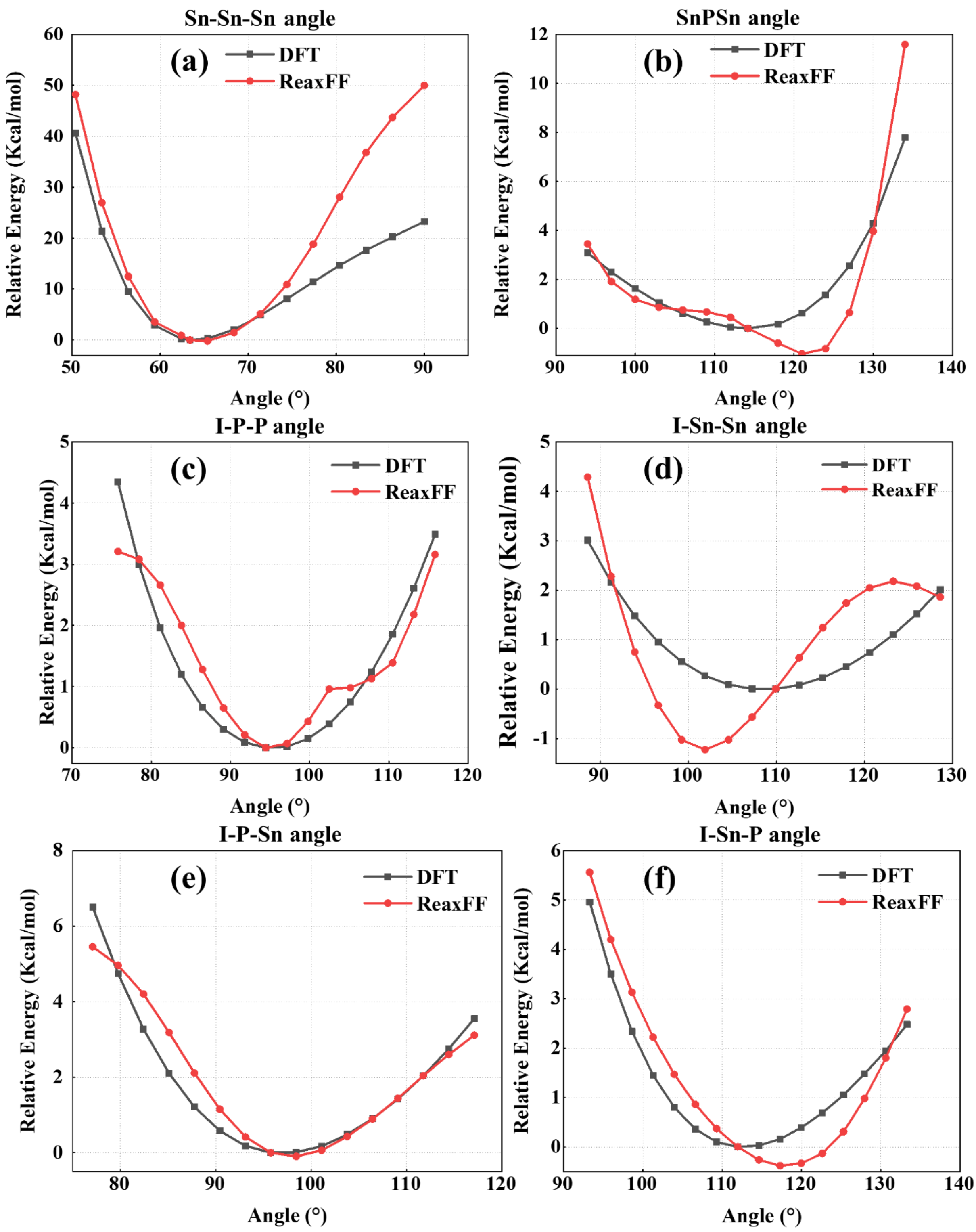

Fig. S3: Comparison between energy calculated from DFT (gray) and ReaxFF (red) for valence angle distortion (a) Sn-Sn-Sn angle, (b) Sn-P-Sn angle, (c) I-P-P angle, (d) I-Sn-Sn angle, (e) I-P-Sn angle, and (f) I-Sn-P angle.

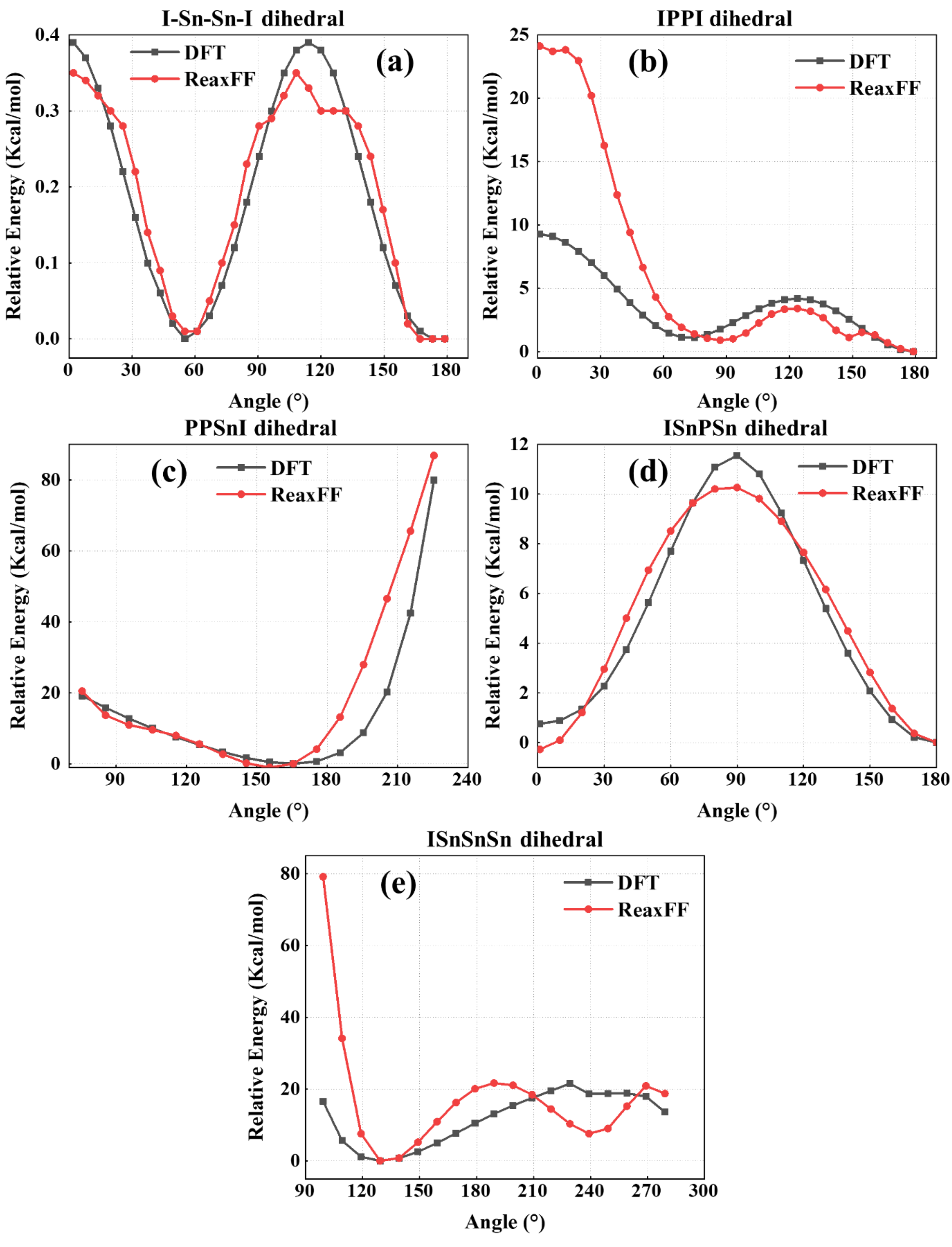

Fig. S4: Comparison between energy calculated from DFT (gray) and ReaxFF (red) for dihedral angle (a) I-Sn-Sn-I, (b) I-P-P-I, (c) P-P-Sn-I, (d) I-Sn-P-Sn, and (e) I-Sn-Sn-Sn angle.

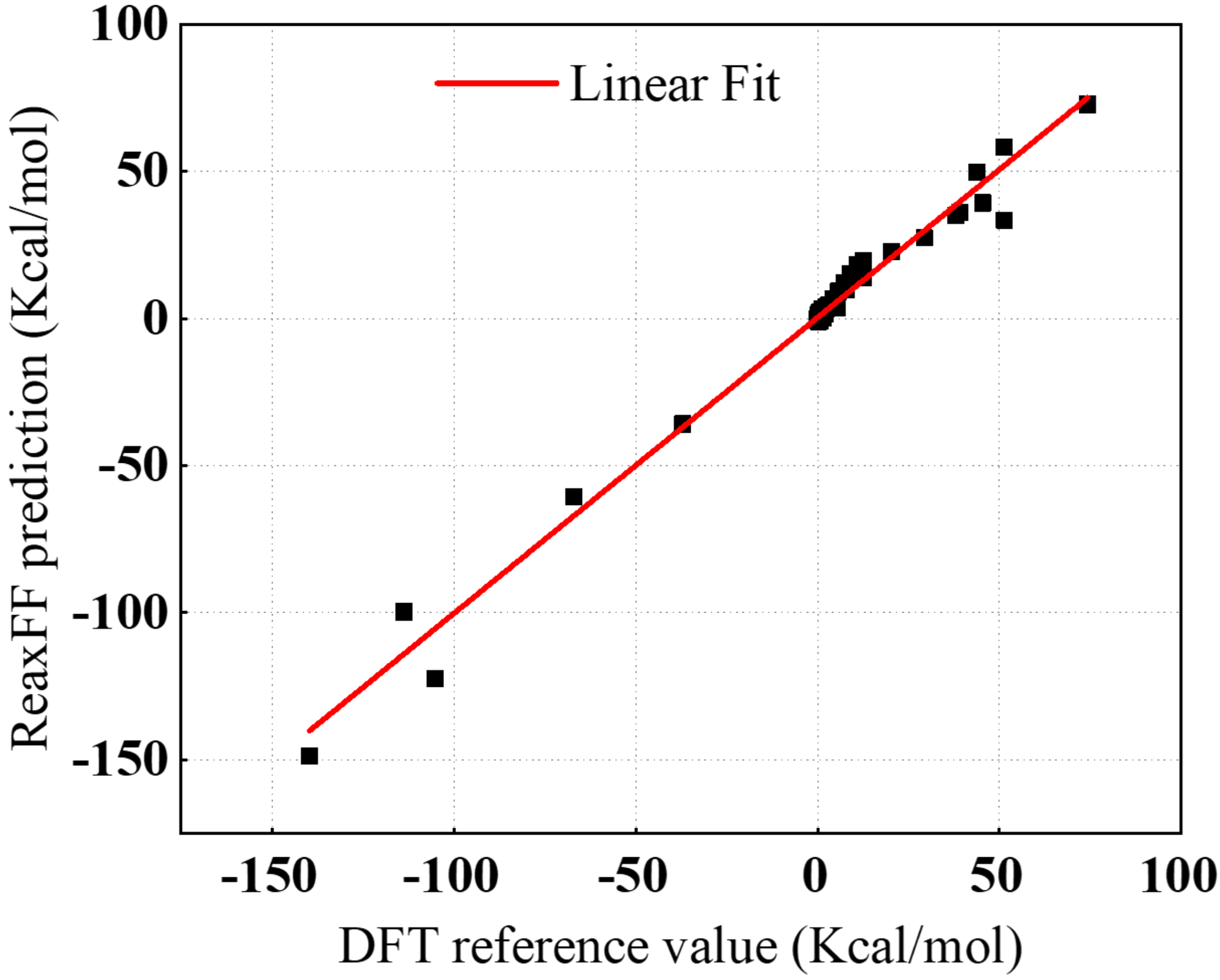

Fig. S5: Quantitative assessment of the agreement between DFT reference values and ReaxFF prediction for α-Sn and β-Sn equations of state, $Sn_x$ (x = 2 – 6) cluster formation energies, and reaction energies. The red line represents the linear fit, which closely follows the y = x relationship.

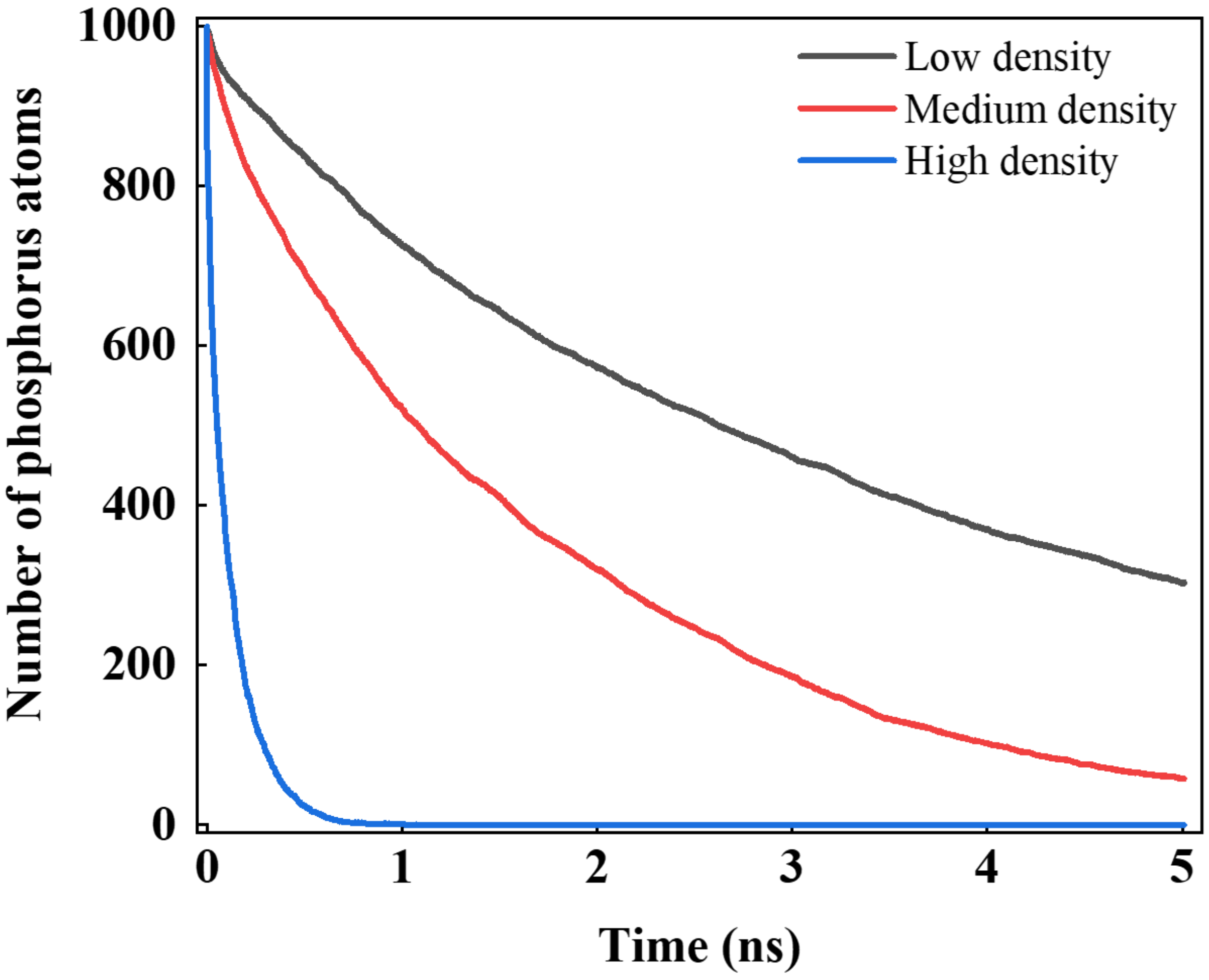

Fig. S6: Phosphorus atom consumption dynamics in P/Sn/$I_2$ system under different densities corresponding to effective pressures of 5 atm, 10 atm, and 100 atm at 923K. The black curve refers to low density, the red curve to medium density, and the blue curve to high density.

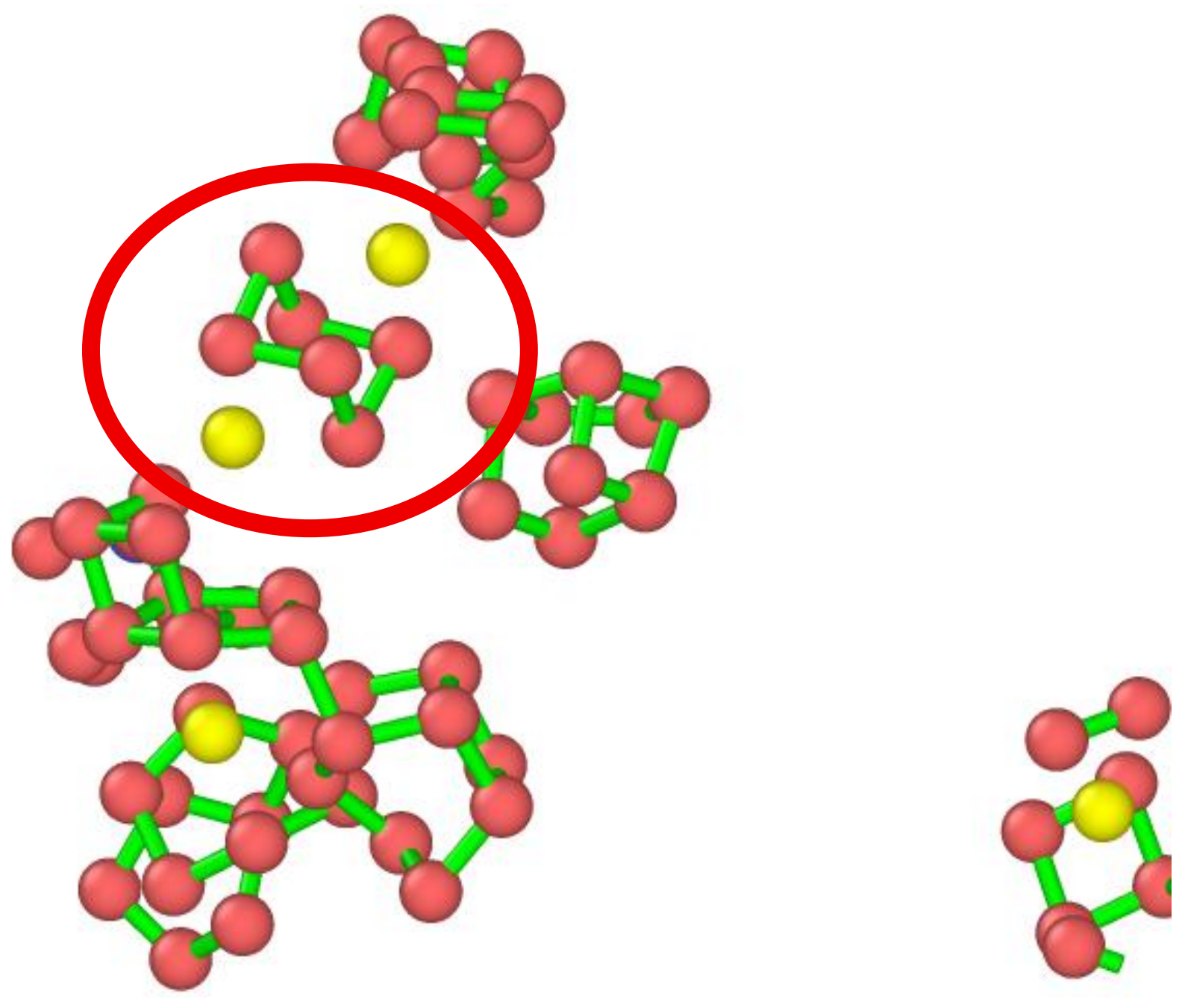

Fig. S7: Snapshot of black phosphorus-like hexagon stabilized by two iodine atoms observed during the simulation.

**Reactive force field parameters for P/Sn/$I_2$ system.**

**Reactive MD-force field: P/Sn/I-2512**

39 ! Number of general parameters

50.0000 !p_boc1 Eq(4c): Overcoordination parameter

9.5469 !p_boc2 Eq(4d): Overcoordination parameter

26.5405 !p_coa2 Eq(15): Valency angle conjugation

1.7224 !p_trip4 Eq(20): Triple bond stabilisation

6.8702 !p_trip3 Eq(20): Triple bond stabilisation

60.4850 !k_c2 Eq(19): C2-correction

1.0588 !p_ovun6 Eq(12): Undercoordination

4.6000 !p_trip2 Eq(20): Triple bond stabilisation

12.1176 !p_ovun7 Eq(12): Undercoordination

13.3056 !p_ovun8 Eq(12): Undercoordination

-70.5044 !p_trip1 Eq(20): Triple bond stabilization

0.0000 !Lower Taper-radius (must be 0)

10.0000 !R_cut Eq(21): Upper Taper-radius

2.8793 !p_fe1 Eq(6a): Fe dimer correction

33.8667 !p_val6 Eq(13c): Valency undercoordination

6.0891 !p_lp1 Eq(8): Lone pair param

1.0563 !p_val9 Eq(13f): Valency angle exponent

2.0384 !p_val10 Eq(13g): Valency angle parameter

6.1431 !p_fe2 Eq(6a): Fe dimer correction

6.9290 !p_pen2 Eq(14a): Double bond/angle param

0.3989 !p_pen3 Eq(14a): Double bond/angle param

3.9954 !p_pen4 Eq(14a): Double bond/angle param

-2.4837 !p_fe3 Eq(6a): Fe dimer correction

5.7796 !p_tor2 Eq(16b): Torsion/BO parameter

10.0000 !p_tor3 Eq(16c): Torsion overcoordination

1.9487 !p_tor4 Eq(16c): Torsion overcoordination

-1.2327 !p_elho Eq(26a): electron-hole interaction

2.1645 !p_cot2 Eq(17b): Conjugation if tors13=0

1.5591 !p_vdW1 Eq(23b): vdWaals shielding

0.1000 !Cutoff for bond order (*100)

2.1365 !p_coa4 Eq(15): Valency angle conjugation

0.6991 !p_ovun4 Eq(11b): Over/Undercoordination

50.0000 !p_ovun3 Eq(11b): Over/Undercoordination

1.8512 !p_val8 Eq(13d): Valency/lone pair param

0.5000 !X_soft Eq(25): ACKS2 softness for X_ij

20.0000 !d Eq(23d): Scale factor in lg-dispersion

5.0000 !p_val Eq(27): Gauss exponent for electrons

0.0000 !1 Eq(13e): disable undecoord in val angle

2.6962 !p_coa3 Eq(15): Valency angle conjugation

3 ! Nr of atoms; cov.r; valency;a.m;Rvdw;Evdw;gammaEEM;cov.r2;#

alfa;gammavdW;valency;Eunder;Eover;chiEEM;etaEEM;n.u.

cov r3;Elp;Heat inc.;bo131;bo132;bo133;softcut;n.u.

ov/un;val1;n.u.;val3,vval4

```
P    2.1199   3.0000  30.9738   2.3355   0.0887   0.4060   1.9507   5.0000
     9.5120   7.6148   3.0000  -1.6536  82.5170   6.3467   8.5658   0.0000
     1.8354   0.0000 120.0000  11.8556  15.5783   2.8491  5066.5788   2.1233
    -2.0858   4.8954   1.0338   3.0000   1.6350   2.6552   0.0743  15.5028
Sn   2.5509   4.0000 118.7100   3.1324   0.0915   0.9297  -1.0000   4.0000
    10.1488   5.7200   4.0000   5.5731   0.0000   4.0471   4.4366   0.0000
    -1.0000   0.0000 133.1770   5.8983   1.4217   4.4778   0.8563   0.0000
   -10.7785   6.5712   1.0338   4.0000   6.6295   2.4059   5.8229  13.3759
I    2.5597   1.0000 126.9000   2.9629   0.1248   0.8535  -1.0000   7.0000
     9.7338   8.9639   1.0000   0.0000   0.0000   9.5847   8.8078   2.0000
    -1.0000   2.3407  35.1770   6.2293   5.2293   0.1542   0.8563   0.0000
   -10.6970   2.9867   1.0338   1.0000   2.5791   2.5211  11.0637  12.4885
```

6 ! Nr of bonds; Edis1;LPpen;n.u.;pbe1;pbo5;13corr;pbo6

pbe2;pbo3;pbo4;n.u.;pbo1;pbo2;ovcorr

1 1 52.2711 23.4911 20.0346 0.4917 -0.2395 1.0000 17.8190 0.7412

1.4218 -0.2226 13.6705 1.0000 -0.2457 7.5884 1.0000 0.0000

1 2 93.3022 0.0000 0.0000 -0.1602 -0.2000 1.0000 16.0000 0.3220

5.1075 -0.2000 15.0000 1.0000 -0.1805 5.8529 0.0000 0.0000

3 1 64.6103 0.0000 0.0000 0.4535 -0.2000 1.0000 16.0000 0.2056

20.4394 -0.2000 15.0000 1.0000 -0.1277 5.4468 0.0000 0.0000

2 2 57.3448 0.0000 0.0000 0.5952 -0.2000 0.0000 16.0000 0.5757

1.9073 -0.2000 15.0000 1.0000 -0.3517 5.3745 0.0000 0.0000

3 2 80.4702 0.0000 0.0000 0.2093 -0.2000 0.0000 16.0000 0.1598

17.8964 -0.2000 15.0000 1.0000 -0.1614 5.4013 0.0000 0.0000

3 3 96.9583 0.0000 0.2561 0.1893 -0.3500 0.0000 25.0000 0.9258

-0.2005 -0.2500 15.0000 1.0000 -0.1576 5.9358 0.0000 0.0000

3 ! Nr of off-diagonal terms; Ediss;Ro;gamma;rsigma;rpi;rpi2

1 2 0.1405 2.8873 10.7239 2.4064 -1.0000 -1.0000

3 2 0.0811 3.2106 10.4947 2.4779 -1.0000 -1.0000

3 1 0.0794 2.5730 10.7545 2.4305 -1.0000 -1.0000

16 ! Nr of angles;at1;at2;at3;Thetao,o;ka;kb;pv1;pv2

1 1 1 81.1291 81.4496 0.5051 0.0000 0.1993 0.0000 1.0534

1 1 2 112.9855 18.6788 7.1657 0.0000 2.1000 0.0000 1.0401

2 1 2 86.9959 7.6767 6.7258 0.0000 1.6323 0.0000 1.3031

3 2 3 78.0469 3.3027 3.8743 0.0000 1.7995 0.0000 3.1607

3 1 3 70.1720 13.7073 3.6987 0.0000 3.6193 0.0000 3.8403

3 1 1 81.5826 20.9186 3.0235 0.0000 2.0845 0.0000 1.0773

2 2 2 74.6120 5.6549 2.6302 0.0000 1.4051 0.0000 2.6688

3 2 2 87.6645 4.5004 5.5847 0.0000 2.4788 0.0000 1.5501

3 2 1 63.2640 24.2394 2.3783 0.0000 0.1427 0.0000 2.1600

3 1 2 64.5602 9.8374 2.5602 0.0000 1.1221 0.0000 1.7464

```
  1  3  1   0.0000  10.0000   1.0000   0.0000   1.0000   0.0000   1.4000
  2  3  2   0.0172   1.5949   4.6751   0.0000   0.1157   0.0000   3.2742
  3  3  1   0.0000   0.5000   1.0000   0.0000   1.0000   0.0000   1.0000
  1  3  2   0.0000   0.5000   1.0000   0.0000   1.0000   0.0000   1.0000
  3  3  2   0.0000  25.0000   1.5000   0.0000   1.0000   0.0000   1.4000
  3  3  3   0.0000  37.9277   8.0000   0.0000   1.0000   0.0000   1.0000
7 ! Nr of torsions;at1;at2;at3;at4;;V1;V2;V3;V2(BO);vconj;n.u;n
3  1  1  3 -16.2095   0.6475  -1.0000 -24.9776   0.0000   0.0000   0.0000
  3  2  2  3  -1.8128   7.6244  -0.3018 -14.5074   0.0000   0.0000   0.0000
  3  2  1  1   0.4955  42.3210  -0.5326  -8.5761   0.0000   0.0000   0.0000
  3  2  2  2 -15.7484   9.9020   0.7023  -0.8383   0.0000   0.0000   0.0000
  3  2  1  2   9.7133  12.3823   0.1831  -0.1855   0.0000   0.0000   0.0000
  0  3  3  0   0.0000   0.0000   0.0000   0.0000   0.0000   0.0000   0.0000
  0  2  3  0   0.0000   0.0000   0.0000   0.0000   0.0000   0.0000   0.0000
0 ! Nr of hydrogen bonds;at1;at2;at3;Rhb;Dehb;vhb1
```